\documentclass[sigplan,nonacm]{acmart}

\usepackage{tabularx}
\usepackage{xspace}
\usepackage{xurl}
\usepackage[x11names]{xcolor}     %
\usepackage{multirow}
\usepackage{amsmath}
\usepackage{makecell}
\usepackage{colortbl} %
\usepackage{booktabs}
\usepackage{pifont}
\usepackage{bbding}
\usepackage{adjustbox} %
\usepackage{comment}

\usepackage{subcaption}
\usepackage{tabularx}
\usepackage{graphicx}
\definecolor{gtblue}{RGB}{0,20,115}

\usepackage[capitalize]{cleveref}
\crefformat{section}{\S#2#1#3} %
\crefformat{subsection}{\S#2#1#3}
\crefformat{subsubsection}{\S#2#1#3}

\crefname{figure}{Fig.}{Figs.}
\Crefname{figure}{Fig.}{Figs.}
\crefname{equation}{Eqn.}{Eqns.}
\Crefname{equation}{Eqn.}{Eqns.}
\crefname{section}{\S}{\S}
\Crefname{section}{\S}{\S}
\crefname{lstlisting}{Listing}{Listings}
\Crefname{lstlisting}{Listing}{Listings}
\crefname{table}{Tab.}{Tabs.}
\Crefname{table}{Tab.}{Tabs.}

\newcommand{\insertFigure}[5]{
    \begin{figure}[t]
      \centering
      \includegraphics[width=#3\linewidth]{figure/#1}
      \vspace{#4}
      \caption{\small #2}
      \label{fig:#1}
      \vspace{#5}
    \end{figure}
}

\newcommand{\insertSubFigs}[9]{
  \begin{figure}[t]
    \centering
    \begin{subfigure}[t]{0.48\columnwidth}
      \centering
      \includegraphics[width=\linewidth]{figure/#1}
      \caption{\centering #2}
      \label{fig:#3}
    \end{subfigure}
    \hfill
    \begin{subfigure}[t]{0.48\columnwidth}
      \centering
      \includegraphics[width=\linewidth]{figure/#4}
      \caption{\centering #5}
      \label{fig:#6}
    \end{subfigure}
    \caption{\small #7}
    \label{fig:#8}
    \vspace{#9}
  \end{figure}
}

\newcommand{\squishlist}{
 \begin{list}{$\bullet$}
  { \setlength{\itemsep}{0pt}
     \setlength{\parsep}{3pt}
     \setlength{\topsep}{3pt}
     \setlength{\partopsep}{0pt}
     \setlength{\leftmargin}{1.5em}
     \setlength{\labelwidth}{1em}
     \setlength{\labelsep}{0.5em} } }

\newcommand{\squishend}{
  \end{list}  }

\newcommand{\cmark}{\CheckmarkBold}          %
\newcommand{\gcmark}{\textcolor{Green3}{\CheckmarkBold}}          %
\newcommand{\xmark}{\ding{55}}           %

\newcommand{\sys}{\textsc{Flint}\xspace}

\begin{document}
\title{Flint: Compiler Enabled Cluster-Free Design Space Exploration for Distributed ML}

\begin{abstract}
Design space exploration for future distributed Machine Learning systems suffers from a lack of readily available workload representation that enables flexible exploration across the stack. We present \textsc{Flint}, a framework that bridges this gap by leveraging the Intermediate Representation of Machine Learning framework compilers. The compiler does the heavy weight lifting of understanding and preserving the behavior of the original model code. \textsc{Flint} can collect the workload representation of arbitrary cluster size because it interfaces with the compiler before hardware execution. We validate the workload graph against post-execution traces and show the flexibility of \textsc{Flint} through a design space exploration case study.
\end{abstract}
\author{Jinsun Yoo}
\affiliation{
\institution{Georgia Institute of Technology}
\city{Atlanta}
\state{Georgia}
\country{USA}
}
\email{jinsun@gatech.edu}

\author{Meghan Cowan}
\affiliation{
\institution{NVIDIA}
\city{Santa Clara}
\state{California}
\country{USA}
}
\email{mcowan@nvidia.com}

\author{Zheng Du}
\affiliation{
\institution{Georgia Institute of Technology}
\city{Atlanta}
\state{Georgia}
\country{USA}
}
\email{zdu@gatech.edu}

\author{Changhai Man}
\affiliation{
\institution{Georgia Institute of Technology}
\city{Atlanta}
\state{Georgia}
\country{USA}
}
\email{cman8@gatech.edu}

\author{Srinivas Sridharan}
\affiliation{
\institution{NVIDIA}
\city{Santa Clara}
\state{California}
\country{USA}
}
\email{srisridharan@nvidia.com}

\author{Tushar Krishna}
\affiliation{
\institution{Georgia Institute of Technology}
\city{Atlanta}
\state{Georgia}
\country{USA}
}
\email{tushar@ece.gatech.edu}

\maketitle %

\section{Introduction}
\label{sec:intro}

Artificial Intelligence (AI) is pervasively influencing our everyday lives through applications such as query engines, image and video generation, and code completion. Recent advancements of AI have been steered by Large Language Models (LLMs), such as Llama~\cite{grattafiori2024llama3herdmodels}, GPT~\cite{openai2024gpt4technicalreport}, or Deepseek~\cite{deepseekai2025deepseekr1incentivizingreasoningcapability}. These models are so large that executing them on a single Graphical Processing Unit (GPU) is infeasible. For example, the recently released LLama4 Behemoth model has 288 billion active parameters, which is too large to fit on a single NVIDIA GPU. As a result, LLM workloads are usually distributed across a large number of GPUs, with the largest clusters extending up to tens of thousands of GPUs. 

\insertFigure{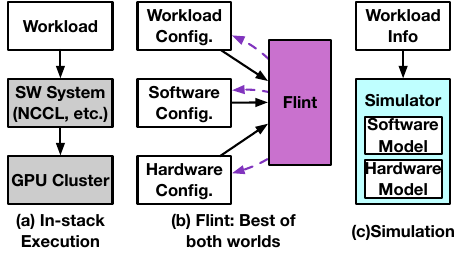}{Different approaches to design space exploration. (a) In-stack execution on real cluster. Users cannot easily study alternate, novel cluster or software system configurations (colored in gray). (b) \sys{} receives configurations across multiple layers and provides feedback, guiding the  configuration search across all areas (purple dashed arrow). (c) Simulations have the best freedom in navigating new configurations but require users to feed workload information.}{1}{-1.5em}{0em}

How do you find the optimal configuration to deploy Machine Learning~(ML) workloads on a future system? 
Design knobs span areas such as parallelization, scheduling, collectives and physical topology. These options form an explosive search space, where the best choice for a job may be suboptimal for another~\cite{maya}.

\begin{table*}[t]
\caption{\small{Comparison of \sys against prior art.}}
\label{tab:comparison}
\vspace{-0.5em}
\begin{center}
\begin{small}
\resizebox{\textwidth}{!}{
\begin{tabular}{l|cc|cccc}
\specialrule{1.2pt}{0pt}{0pt} %
 & \multicolumn{2}{c|}{\textbf{Workload Graph}}     & \multicolumn{4}{c}{\textbf{Design Space}} \\
 & Cluster-Free & Source Code & Scheduling & Parallelization & Custom Collective & Topology \\ %
\midrule
Post-execution Traces + Simulation \cite{astrasim-2.0}      & \xmark & \xmark & \cmark & \cmark & \xmark & \xmark   \\ %
Synthetic Generation + Simulation \cite{calculon, stage}  & \cmark & \xmark & \cmark & \cmark & \xmark & \cmark   \\ %
Runtime Compilers \cite{pytorch2} & \xmark & \cmark & \cmark & $\triangle$ & \xmark & \xmark   \\ %
CUDA API Capture \cite{maya,phantora}   & \cmark & \cmark & \xmark & \cmark & \xmark & \cmark \\
\textbf{\sys (Our work)}  & \gcmark & \gcmark & \gcmark & \gcmark & \gcmark & \gcmark   \\ %
\specialrule{1.2pt}{0pt}{0pt} %
\end{tabular}
}
\end{small}
\end{center}
\end{table*}

One approach is to simply run the whole software stack on real systems and rank the performance of different configurations. However, in-stack executions are constrained to the provided system configurations, limiting the design space. Users simply cannot deploy the software on future systems of unprecedented scale, new hardware, or a different connectivity. Even existing largescale systems are in high demand making it costly for researchers (in academia) or are prioritized by production teams (in industry). When access \textit{is} granted, the limits of existing software stack restricts exploring alternate optimizations (\cref{subsec:dse}).

In the other extreme lies cost model (i.e. simulator or emulator) based frameworks. While this approach can search through several hypothetical configurations, deciding which workload information to model remains a challenge. Running a workload on a real GPU cluster and obtaining \textit{post-execution} requires the same number of GPUs as an in-stack execution, diminishing the benefit of cost models~\cite{chakra, astrasim-2.0}. Synthetically generating a \textit{pre-execution} representation from a symbolic workload description does not need GPU, but developers must manually extend the synthetic generator with each new workload optimization. Orthogonally, some frameworks build their own pipeline to generate workload information~\cite{maya,simai}. However, these pipelines are constrained to their own cost model, hence not leveraging the full set of available cost models (\cref{sec:background}).

We propose \sys, a novel framework that aims to bridge the best of both worlds\footnote{The name is inspired by how in ancient times flint made it much easier to start fires, granting more time for usecases such as cooking, lighting, or keeping the house warm.}. The key approach of \sys is to leverage the compiler Intermediate Representation (IR) from the ML framework's compiler (for example, FX Graph from PyTorch) and feed it to cost models~\citep{fx, pytorch2}. The compiler IR preserves the behavior of the source code, including the true data dependencies between operators. Users do not have to configure external synthetic generators, but simply reuse already available model code. In addition to the benefit of easy workload graph generation, \sys{} feeds the workload graph to a variety of cost models to showcase a versatile set of usecases. To do this, it converts the captured IR into a Chakra graph~\cite{chakra}. Chakra is a standard ecosystem maintained by MLCommons and widely used across several companies~\citep{keysight_kai_datacenter_builder, feng2023chakra_blog}. The ecosystem is centered around the Chakra graph, which is used across multiple downstream tools such as simulators or replay tools. Examples include the ASTRA-sim simulator, Genie network emulator, and many proprietary cost models that leverage Chakra graphs~\citep{keysight_kai_datacenter_builder, genie, astrasim-2.0}.

There are two challenges to enabling \sys{}. First, \sys{} has to extract the workload graph from the right level of IR. ML frameworks, such as PyTorch, perform several iterations of optimizations and lowering on the IR once it is obtained from the workload graph. There is a tradeoff between forcing a platform specific optimization and restricting the search space, and being too high level and generic. We provide a deep discussion and analysis of the available compiler tradeoffs, and argue for the right stage of FX graphs (\cref{sec:design}). 

Second, \sys{} needs to provide the illusion to the ML framework that it is running on a real GPU cluster. A key design principle of \sys{} is for the developer to `run' the model with the source code as-is, with minimal modifications. If this is not done correctly, the ML framework might attempt to allocate GPU memory or capture an unrealistic IR graph thinking it is running on a CPU only machine. (\cref{sec:runtime}) 

This work provides the following contributions. 
\squishlist
    \item We present \sys{}, the first system to capture and leverage workload information without GPU clusters from compiler IR.
    \item We provide an in-depth discussion on existing workload optimization and the right level of abstraction to capture workload graphs.
    \item We showcase how \sys{} addresses a number of usecases across the stack by leveraging the workload graph across a set of diverse downstream tools.
\squishend

The remainder of the paper is structured as follows: \cref{sec:background} discusses the relevant background to this work. \cref{sec:design} establishes our design principle and explains the design choices we make. \cref{sec:runtime} describes how \sys{} is implemented and the considerations taken for it to work smoothly. \cref{sec:validation} validates the graphs \sys{} generates, and \cref{sec:eval} showcases several usecases that benefit from \sys{}. \cref{sec:discussion} discusses additional issues and outlines future work.

\section{Background and Motivation}

\label{sec:background}

\subsection{Design Space Exploration Overview}
\label{subsec:dse}

\insertFigure{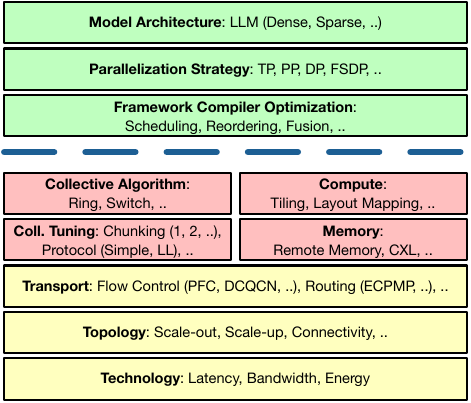}{The layers in distributed ML and their design choices. The boxes correspond to workload (green, top), software system (red, middle) and hardware system (yellow, bottom) related options. The blue dashed line separates workload related options and system related options.}{1}{-1.5em}{-1.5em}

\cref{fig:dse} shows the multiple layers of design choices involved in largescale AI/ML and possible options. In DSE, the goal is to find the best combination of these options that yields the best desired metrics and constraints, such as end to end duration, memory usage, or power consumption. This can be understood as a two step process, divided by the dashed blue line in the figure. First, the developer prepares a description of the workload, either by writing model code from scratch or using synthetic generators (\cref{subsec:graph}). Then, the developer `observes' how the workload performs on a given system, either by executing on a real system or running cost models such as simulators or emulators (\cref{subsec:cost_model}). A good end-to-end DSE framework should enable users to easily explore all of the layers or focus on any specific layer, depending on the usecase.

\subsection{Obtaining Workload Representation}
\label{subsec:graph}

The start of DSE is to understand what the workload looks like. Graph based representations, where each operations and their dependencies are recorded as vertices and edges, are commonly used in prior art. Example usage includes describing a novel model structure, illustrating how a parallelization strategy introduces new compute and communication patterns, or how operations can be reordered for better performance while respecting the workload dependency. Examples include Chakra or GOAL~\cite{chakra, goal}.

However, obtaining these graph representations is a difficult task. A simple approach is to synthetically generate graphs from a symbolic description. A symbolic description lists factors such as the number of layers and size of the hidden dimension. This approach is based on the largely repetitive nature of the transformer model. However, such generic representation is limited in capturing the unique characteristics of different model architectures. For example, Deepseek has the same high-level transformer structure but replaces MHA with multi-latent attention and FFN with Mixture of Expert (MoE) layers. Using SwiGLU was a distinguishing factor of early Llama models. 

Additionally, parallelization strategies make it difficult for synthetic graph generators to keep up with fast moving trends. New parallelization strategies introduce changes to the workload graph that are different from past strategies. Even within the same parallelization strategy different flavors, such as DDP Optimizer which buckets small AllReduces in Data Parallel into a smaller number of larger collectives, add additional burden. While it is possible to extend symbolic representations to represent the different modifications listed above, manual effort is needed to implement and validate each new model or parallelization optimization.

Another approach is to execute the model code on real GPUs and trace which operations were executed. Chakra, for example, supports a workflow to deploy the workload on GPU clusters and collect Chakra Execution Traces (Chakra ET). Here users can easily shift between model structures and parallelization strategies already implemented in the source code without having to reinvent the wheel in synthetic generators. However, this restricts the user to whichever GPU cluster is available.

\insertFigure{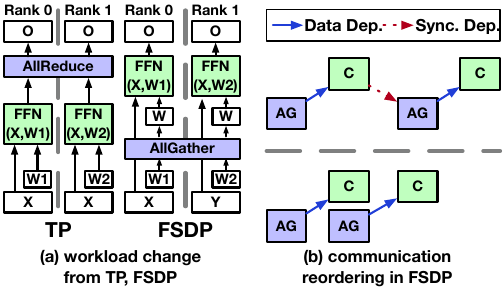}{Various changes in a workload graph. (a) Tensor Parallel and Fully Sharded Data Parallel in a transformer model. W1,W2: partial weights, W: full weight. FFN: Feed Forward Network. X, Y: different input. (b) Scheduling strategies on FSDP. (Top): Synchronization dependency to delay AllGather and save memory. (Bottom): Reordering to maximize compute and communication overlap.}{1}{0em}{0em}

Some works execute the source code as-is, but intercept calls at the CUDA API level and reconstruct the workload graph without actually executing on a real GPU cluster~\cite{maya, phantora}. However, these approaches contain false dependencies, restricting them from exploring graph modifications such as scheduling optimizations. This is because these work only have access to the low level hardware API. These work infer the dependency from the order of API calls or synchronization events that are manually injected. Such inferred dependencies do not represent the \textit{true} data dependency between operations. 

Take the example depicted in \cref{fig:graph}b, which describes the collective reordering introduced in SimpleFSDP~\cite{simplefsdp}. While compute operations in FSDP are dependent on the AllGather collectives to gather the model weights (data dependency), the AllGather collectives themselves are not dependent on any prior operation. However, the original implementation of FSDP injected a synchronization dependency between the AllGather and compute operations of the previous layer, exposing the AllGather communication to delay the Gather and limit active memory. The reordering strategy explores the tradeoff between memory and latency by running later collectives upfront and overlapping them with earlier compute or waiting to run collectives until the weights are needed by the compute. Because the CUDA API based approaches rely on synchronization information, they cannot tell if the communication can be issued earlier.

The above discussion highlights a gap in the ability to capture comprehensive workload representations without needing large number of GPUs. While in theory any graph can be converted to use the Chakra schema, the key distinction that sets \sys{} apart is where the graph originates from - the compiler IR.

\subsection{Cost Models and Usecases}
\label{subsec:cost_model}

Cost models help users understand systems that they do not have full access to. Because there are multiple usecases, one-size-fits all does not apply. 

\textbf{Simulators} Simulators help model hypothetical systems in both software and hardware. In software, simulators can model arbitrary collective implementations beyond the standard Ring and Tree implementations. Simulators can also easily model hardware topologies, whether they involve a large number of ranks, complex interconnect connectivity, or novel technologies such as  new transports, wafer scale substrates, optical interconnects. Examples include ASTRA-sim, ATLAHS, the simulator in SimAI, etc~\cite{astrasim-2.0, atlahs, simai}.

\textbf{Emulators} Emulators model the system to some extent. For example, when studying the behavior of network fabrics such as NICs or Switches, provisioning all GPUs is not necessary. Emulators that generate traffic based on the workload pattern allows users to study real network behavior and features without relying on system implementations. Examples include Genie or Keysight AI Datacenter Builder (KAIDCB)~\citep{keysight_kai_datacenter_builder}.

The cost models report metrics such as job completion time, memory consumption, or network flow. These metrics guide the next choice on workload, software, or hardware configurations. This feedback loop is repeated until an optimal configuration is found. 

The level of abstraction of the workload representation dictate how much freedom the cost model has in exploring different configurations. %
Recall the graph reordering example discussed in \cref{subsec:graph}. If the workload is able to capture the true data dependency, frameworks can search through different scheduling and fusing strategies by modifying the graph while respecting the intended dependency.

The context where the cost model is running also determines the flexibility. For runtime compilers that leverage pre-execution graphs, they can only search through graph modifications, but not all of the other spaces above, especially parallelization. This is because they are fixed as a runtime component, and all of the other factors are already discussed by the system and are out of their scope.

Cost models are well studied in prior art. However, they were previously limited due to the shortcomings on how the workload graphs are generated (\cref{subsec:graph}). \sys{}'s approach of leveraging compiler IR greatly enhances the search space of these cost models. %

\subsection{Model Compilation and Execution in PyTorch}
\label{subsec:pytorch}

\insertFigure{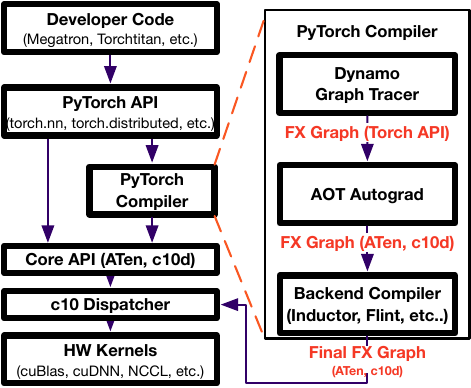}{The PyTorch software stack and the PyTorch compiler.}{0.8}{-1em}{-1em}

\begin{table}[t]
\caption{\small Example Graph Passes in the Compilation Process.}
\label{tab:passes}
\vspace{-0.5em}
\centering
\begin{small}
\begin{tabular}{ll}
\toprule
\textbf{Compilation Stage} & \textbf{Graph Passes} \\
\midrule
Pre AOTAutograd & Fuse matmul \& permute into\\ & transpose \& matmul, etc.  \\
Post AOTAutograd & Remove noop, redundant views, etc. \\
\midrule
\multicolumn{2}{c}{\textit{This is where Flint captures the FX Graph}} \\
\midrule
Backend Compiler & FSDP AllGather reordering,\\
                 & DP AllReduce bucketing,  \\
                 & TP micro pipelining \\
\bottomrule
\end{tabular}
\vspace{-1em}
\end{small}
\end{table}

\cref{fig:pytorch} depicts PyTorch's software stack at a high level. Developers define the model structure or apply parallelization by writing code with the PyTorch API or libraries such as Megatron or Torchtitan. These high level functions are broken down into \texttt{aten} or \texttt{c10d} functions for compute and communication operations, respectively. PyTorch executes this code in two ways. In the first approach, \texttt{eager}, the c10 Dispatcher simply runs the code one function at a time. Because it does not have a view of the whole program, PyTorch cannot make global optimizations.

The other option uses a Just in Time (JIT) compiler to capture the whole workload into an Intermediate Representation (IR) called FX Graph.
\cref{fig:convert} shows a sample PyTorch code and the corresponding FX Graph. Each node represents an input tensor variable or a function that generates an intermediate tensor. A node contains metadata of the tensor it represents such as the shape, and pointers to upstream nodes corresponding to the arguments of the function that produced this tensor.

The right part of \cref{fig:pytorch} details the compilation process. Dynamo, PyTorch's frontend compiler, symbolically traces the PyTorch code by extending the Python runtime to record function calls instead of actually executing them. Ahead of Time Autograd (AOT Autograd) then lowers this into \texttt{aten} or \texttt{c10d} level and also creates a graph for the backward pass.
The FX Graph is passed to a backend compiler through a publicly supported API. The backend compiler generates the final FX graph, which is executed by the Dispatcher. 

Throughout this process, Dynamo, AOTAutograd, and the backend compiler makes \textit{graph passes} that modify the captured graph before passing to the next stage. Graph passes add, remove, change, or reorder graph nodes, effectively changing the final operation, while maintaining the proper behavior.
\cref{tab:passes} groups the graph passes by the stage where these passes occur. The passes in the first two stages are largely cosmetic passes, such as removing no-operation functions (multiply by 1, add 0, etc.) or rewriting the same operation for better code generation. These passes are relatively straightforward and do not leave much room for changes. PyTorch does not expose options to enable or disable these passes.

On the other hand, the graph passes in the backend compilers are more interesting. The collective reordering example depicted in \cref{fig:graph}(b) is implemented within the default Inductor backend compiler. Here, the optimizations introduce a tradeoff or are not yet fully understood, leaving room for interesting research questions.PyTorch enables a wide range of flexibility by exposing knobs to enable or configure these passes, and even exposes an API that developers can implement to write their own backend compilers with custom graph passes. \sys{} leverages this endpoint to capture the FX Graph already traced by PyTorch's compiler (\cref{sec:design}).

\begin{figure}[t]
  \centering
  \includegraphics[width=0.8\linewidth]{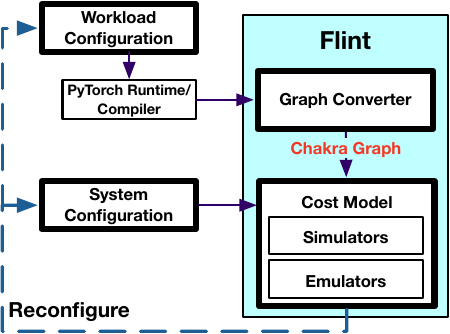}
  \caption{\small High level depiction of \sys{} architecture. Developers provide workload configuration (i.e. PyTorch code) and the system configuration. The workload code is captured by the PyTorch compiler into an FX Graph, which \sys{}'s Graph Converter converts into a Chakra Graph. The system configuration configures the cost model. The cost model generates metrics, which are used to select the next set of configurations (blue dashed arrows)}
  \label{fig:arch}
  \vspace{-1 em}
\end{figure}

\section{System Design}
\label{sec:design}

\subsection{Design Principles}
\label{subsec:principle}
We first discuss the design principles behind \sys{}.

\textbf{P1: Generate a graph compatible with existing multiple cost models: } 
\sys{} should provide an easy way to obtain workload graphs that can be fed into cost models. We do not wish to tie \sys{} to a specific cost model.

\textbf{P2: Capture the source code behavior: }
We want to correctly capture the unique behavior of model architectures, parallelization strategies, and their implementations (\cref{subsec:graph}). \sys{} should simply take the source code and generate the graph, instead of having to manually write the logic to synthetically create graphs from symbolic descriptions. The operations should capture necessary information to recreate the operation, such as the shape of input and output tensors.

\textbf{P3: Balance between relevant detail and freedom of exploration: }

Capturing workload information closer to the hardware execution ensures the information closely represents existing hardware and software stack, but does not leave room for different configuration options or futuristic optimizations.

On the other hand, abstracting away components with little ambiguity yields no significant benefit. For example, PyTorch internals such as the decomposition of Torch IR into \texttt{aten}/\texttt{c10d} operations, is a framework implementation issue and is beyond the scope of works that P1 targets.

\textbf{P4: Easily usable with little hardware demand: }
\sys{} should be easy to use. Not requiring multiple GPUs is the cornerstone to achieving this goal. Specifically, users should not need to build physical clusters for every system configuration they want to model. 

\subsection{Design Choices Behind \sys{}}
\begin{center}
\fbox{\begin{minipage}{0.96\linewidth}
\sys{} uses a \textit{custom backend compiler} to capture the \textit{FXGraph after AOT Autograd}, and generates \textit{pre-execution} graphs using the \textit{Chakra schema}. It then feeds this graph to a set of cost models depending on the usecase.
\end{minipage}}
\end{center}

\cref{fig:arch} depicts the workflow in \sys{}. The developer first registers \sys{} as a backend compiler to PyTorch.
Then the user defines a set of model configurations such as the architecture or parallelization strategy, and writes PyTorch code. The compiler's frontend components traces this code and generates an FX Graph, which is provided to \sys{}. \sys{} runs a set of graph modification passes that the user picks, and then converts the FX Graph to a Chakra graph ~\cite{pytorch2}.

This is then fed to the cost models, which are configured with system software or hardware configurations (collective algorithm, hardware specifications, etc.). The resulting metrics are used as feedback to choose a new set of configurations.

We now provide our rationale behind this design choice, and how it helps fulfil our design principles. 

We first discuss how our decision to capture from the compiler IR helps us fulfill both P2 and P4. Extracting information from the source code allows us to capture the newest updates to the PyTorch runtime without reinventing the wheel of generating the graph from a symbolic representation.

We note that when \sys{} extracts a graph after a certain stage, it is equivalent to declaring that the developer is not interested in exploring the optimizations available in that stage. When there are design choices the user wants to study, \sys{} should capture the graph \textit{before} the optimization stage is fixed, so that the user can switch between alternate configurations.

In that aspect, we do not capture lower level information such as the collective algorithm is being used (\cref{subsec:dse}). Assume that NCCL would have chosen the Ring algorithm when running a model on an existing cluster. Fixing this information in the workload representation restricts users from searching other algorithms such as Tree or custom non-standard synthesized algorithms.

We also make the design choice of not capturing the FX graphs at the Torch IR level, but wait to capture after AOT autograd. The biggest reason behind this is the backward pass. AOT autograd, unlike Dynamo, can capture the backward pass which is also important in modeling different configurations. Additionally, we find Torch IR to be too high level. PyTorch's internal runtime already breaks down the Torch IR into \texttt{aten}/\texttt{c10d} operations regardless of other configurations. Capturing at the higher Torch IR level leaves room for downstream tools to decompose the operations differently, which results in a workload behavior different from PyTorch code (P3). Finally, we find the graph passes in this stage to be trivial, evidenced by the lack of PyTorch exposed knobs to configure them. Therefore, there is little merit for \sys{} to capture the graph before this stage.

However, we do not capture the backend compiler optimizations. This is an interesting search space and we do not want \sys{} to restrict the user's search space by forcing a single option on the graph. Additionally, such optimizations, if captured, would be tied to the platform. For example, one possible graph transformation is to maximize compute-communication overlap. Because the compute and communication duration differs on the hardware, the reordering result for one platform would not be valid on another platform. Finally, there is no definitive backend compiler to capture. While PyTorch uses Inductor by default, there are other backend compilers such as TensorRT. Capturing the result of different backend compilers would yield different results.

The FX Graph that \sys{} captures right after AOTAutograd, therefore, could be different from what is executed if users run PyTorch (and thus Inductor) out of the box. \sys{} can elect to apply these backend compiler graph passes (or a combination of) before feeding into the cost model. This provides a graph that would happen with default PyTorch execution. This is useful if the developer is not interested in workload optimizations and more interested in system optimizations. On the contrary, a developer interested in workload optimizations would elect to work on the unoptimized FX graph that has only the true dependencies. \sys{} exposes this option as a configuration knob, so that developers can easily choose depending on their needs.

Our choice of generating graphs with the Chakra schema allows seamless integration with existing tools that already use Chakra graphs (P1), most notably ASTRA-sim and later Chakra-based collective-simulation workflows~\citep{astrasim-2.0, genie, collectiveapi, keysight_kai_datacenter_builder}.

Note how \sys{} uses the same \textit{schema} as post-execution traces of real execution, but is pre-execution and does not rely on real clusters to run (P4).

P1 also clarifies why simply using the FX Graph as-is is not enough. To the best of our knowledge, there is no cost model that takes in FX Graph. To use FX graphs with existing tools, we would need to convert and extract attributes from FX graphs anyways. Furthermore, FX graphs are specific to the PyTorch framework. On the contrary, Chakra is a framework-neutral schema that allows developers to capture workload information regardless of the framework~\footnote{While \sys{} currently focuses on FX graphs, we envisioning expanding it to other frameworks such as JAX.} Finally, sharing FX graphs between peer researchers is a challenge that makes using FX graphs as-is infeasible. While \texttt{torch.export} saves the captured graph, it saves a serialized Python object. Additional conversion is required to feed this graph into cost models that are not written in Python.

Finally, using a custom backend compiler makes it easy to develop, use, and maintain \sys{}. The API between PyTorch and custom backend compilers, through which PyTorch provides the captured FX graphs, is publicly exposed and relatively well established. This makes it less susceptible to internal or future changes. \sys{} simply implements this API, receives the FX Graph, creates Chakra graphs, and rus DSE passes.

\section{Flint Runtime}
\label{sec:runtime}
\label{subsec:pytorchfix}

\subsection{Execution Model}
In a traditional in-stack deployment, the developer launches $N$ instances of the same PyTorch program, where $N$ is the number of GPUs. Each process applies the parallelization strategy (separate from the code, likely provided through an external configuration) and decides implementation details such as which other ranks to communicate with or how to break down the compute. 

\sys{} preserves the developer experience by deploying $N$ processes with the same model code. However, the processes are not assigned to a physical GPU. Instead, \texttt{torch.compile} of rank $r$ captures the compiler IR of that rank \textit{before} executing any code on a physical GPU, and passes it to \sys{}. The \sys{} converter within each process independently processes the FX Graph. Because the compiler captures the FX Graph after parallelization is applied to the model, each FX Graph is unique to the rank and preserves details such as what are the peer ranks in a collective.

Note how in the traditional in-stack deployment the number of processes per host machine is capped by the number of physical GPUs on the machine. On the other hand parsing the IR in \sys{} is purely a CPU operation. Users can decide to allocate all processes on a single host, or distribute across multiple hosts, depending on the number of processes and their ranks.

\subsection{Illusion of a GPU Runtime}

A key principle of \sys{} is to allow users to collect the workload graphs using the model code and scripts (\texttt{torchrun, mpirun}, etc.) to run these graphs. To seamless collect FX graphs of a model code in from the compiler, \sys{} needs to provide PyTorch the \textit{illusion} that it is running on a GPU cluster. One important aspect is to prevent PyTorch from running operations that require real GPUs. Here, \sys{} requires users to initialize models on the 'Meta device', instead of a GPU. 'Meta device' a fake device that PyTorch supports, where each tensor is simply a data structure that contains only metadata such as the tensor shape, but does not allocate any data value in device memory. Operations on Fake Tensors produce another Fake Tensor with the resulting tensor shape recorded as metadata. This is to prevent PyTorch from attempting to allocate tens of GB of memory on GPU, which it does not have. 

This is the only modification to user code that \sys{} requires. Since \sys{} does not run actual training, it is acceptable to not allocate or compute any real value. 
However, using Meta devices could become an issue in operator dispatch (\cref{subsec:pytorch}). For example, Scaled Dot Product Attention (SDPA) can be translated into a single fused implementation or broken down into simple primitives such as \texttt{add} or \texttt{multiply}. During the symbolic tracing, the PyTorch API function decides \textit{before reaching the dispatcher} whether to break down SDPA. If the tensor was originally allocated to a GPU, a fused kernel is used, but if the tensor was originally allocated on another device (such as the CPU or the Meta device), the SDPA operation is broken into the primitive version. To prevent this, we modify the PyTorch API function of SDPA to treat tensors as if they were allocated on the GPU. While this approach requires direct modifications to the API code, we believe such modifications will be limited in the future, as the aten api does not change rapidly.

Once the runtime obtains the FX Graph, it converts this to a format that can be used by multiple cost models. \sys{} makes a specific choice to convert the FX Graph into Chakra graphs, but it could be converted into different graph formats based on the usecase. Detailed discussions on the conversion process are presented in~\cref{subsec:convert}.

\subsection{Creating Chakra Graphs from FX Graphs}
\label{subsec:convert}
While PyTorch's compiler uses and provides FX graphs as its IR, to our knowledge, no full-stack cost model uses FX graphs as a workload representation input. Instead, we leverage Chakra. Chakra is similar to FX graphs in that it also uses a graph based representation, while it is more widely accepted as an input to downstream cost models~\cite{chakra,astrasim-2.0,genie}.

\sys{} converts the FX Graph it obtains from PyTorch into a Chakra graph. \cref{fig:convert} depicts how a PyTorch snippet is captured into an FX Graph and converted into a Chakra graph. Note how in Chakra a host side function (\texttt{aten::addmm}) and the launched GPU kernel (\texttt{GPU\_COMP::addmm}) is separated. While FX Graph represents dependency after a collective kernel using \texttt{wait\_tensor}, Chakra reprents it using dependencies to both the CPU and GPU operation. Also note that FX nodes that correspond to an input tensor, and not an operation, such as \texttt{primals\_1}, are ignored in Chakra. Both graphs hold information needed to recreate the operation, such as the tensor shape and data type.

During the conversion \sys{} adds the duration of compute operations to the compute nodes. This information is supplied through an offline profiling. While a cluster scale GPU is difficult to obtain, we believe it is relatively reasonable to gain access to a single GPU to profile compute operations. The downstream cost model may choose to either use these profiled numbers or ignore them and leverage their own compute simulations or emulations. This is needed in cases such as modeling future hardware, or hardware that exists but the user simply does not have access to.

\insertFigure{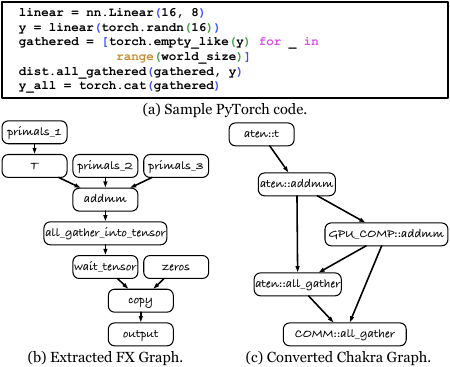}{A sample PyTorch code and the corresponding FX Graph and Chakra Graph.}{1}{-1.5 em}{0em}

\subsection{Cost Model}
\label{subsec:simulation}
Once the Chakra graph is obtained, \sys{} can feed this into a set of cost models. Note that \sys{} is not necessarily restricted to a single cost model. Users can choose between publicly available or proprietary cost models depending on their use cases~\cref{subsec:cost_model}. Depending on the outcome of the cost model, the developer could elect to change the workload configuration, the system configuration, or both. When changing only the system configuration, the developer will reconfigure the cost model but use the same Chakra graph. On the contrary, when changing the workload configuration, the developer will recapture the Chakra graph with the net workload configuration and feed it to the cost model. Leveraging cost models allow users to study systems that are not easily accessible with the simple in-stack execution method.  

\section{Validation}
\label{sec:validation}

We demonstrate the faithfulness of Chakra graphs collected from \sys{} by comparing them against post-execution traces obtained from executing a model.

\subsection {Environment Setup}
We collect Chakra graphs using both \sys{} and post-execution traces. We collect post-execution traces on a physical cluster where each node has 8 NVIDIA H100 GPUs and the nodes are connected through a single 100Gbps InfiniBand HCA. Our model code is based on the Torchtitan framework as of October 19, 2025 \cite{torchtitan}. We use an unmodified PyTorch nightly version \texttt{20251019+cu129} to gather post-execution Chakra traces, while we apply the minimal modifications discussed in \cref{subsec:pytorchfix} to run \sys{}\footnote{Upon acceptance of the paper we will opensource our code and traces as artifacts.}. For the purpose of validation we apply the graph passes that Inductor applies out of the box to the FX graph. We further evaluate usecases that look at the tradeoffs of novel graph passes in ~\cref{subsec:usecase_fsdp}.

\subsection{Operation Count Validation}
\label{subsec:opcount_validation}

\insertFigure{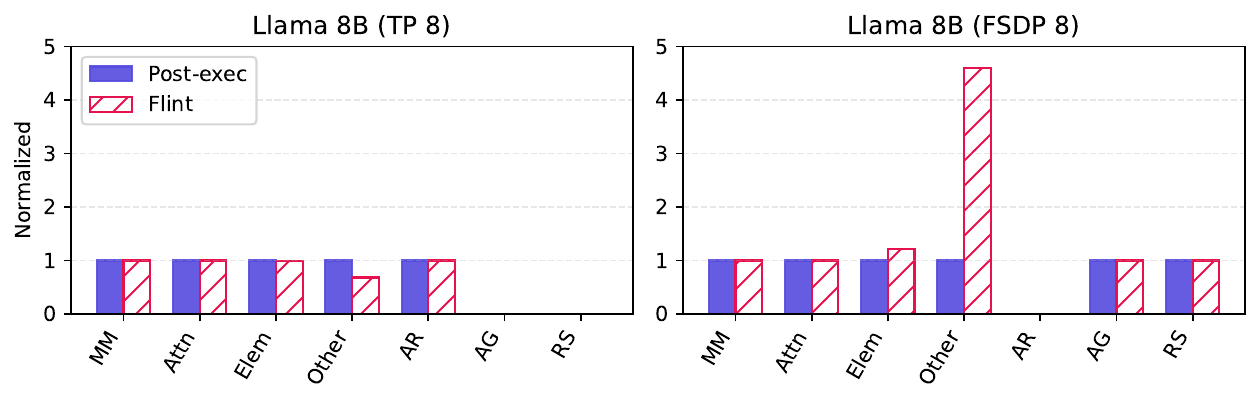}{Counts of operator in \sys{} generated graphs, normalized to post-execution Chakra traces per operator type. MM: GeMM, Attn: Attention, Elem: Elementwise, AR: AllReduce, AG: AllGather, RS: ReduceScatter.}{1}{-2em}{0em}
\insertFigure{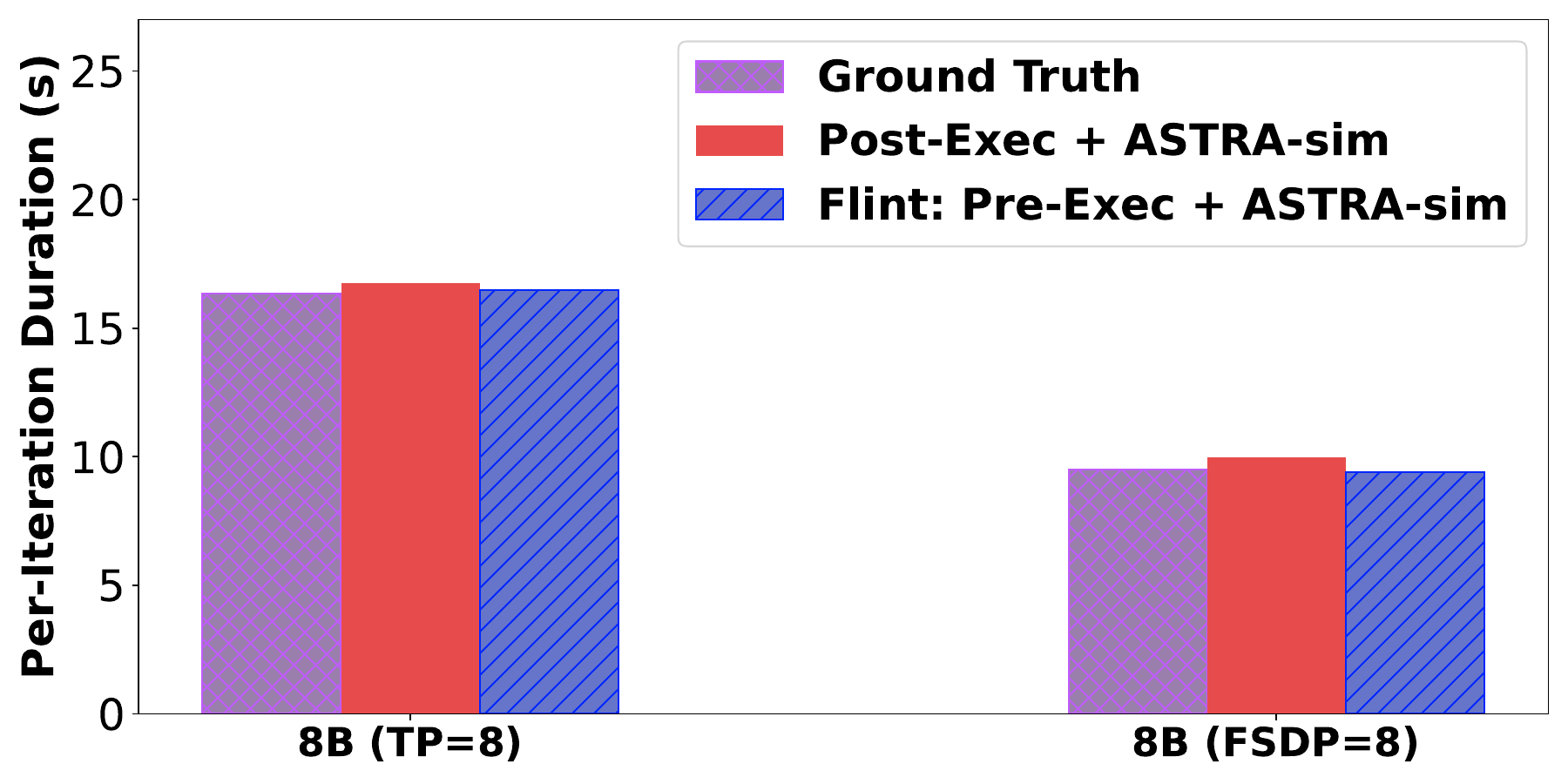}{End to end measurement of per-iteration duration}{1}{-2em}{0em}

\cref{fig:opcount} compares how many times each type of operator occurs in the graphs. Operations that do not occur are marked with a lack of bar. Because the number of operations vary across different types, we normalize each count to the number of occurrences in the post-execution trace. We see that the number between the two sources largely match, especially for GeMM, Attention, and collectives, which are the more important operations.

We note some differences in miscellaneous operations. This is largely due to differences in how \texttt{aten} operations are decomposed into CUDA kernels. For example, one \texttt{aten::mm} operation is decomposed into a cuBLAS kernel and a reduction kernel. These low level decisions are hardware specific and are out of scope of workload graph capture.

\subsection{End to End Validation}

We compare the end to end workload duration of the configurations observed in \cref{subsec:opcount_validation}. Both configurations are issued on a single node and communication occurs over the scale-up network.

We measure the end to end duration with three methods. Ground Truth is the duration when running PyTorch on real GPU clusters. Post-Exec + ASTRA-sim takes the post-execution trace obtained from the Ground Truth run, and feeds it to the ASTRA-sim simulator. Finally, Flint uses the pre-execution graph obtained from PyTorch's FX Graph, and feeds it into ASTRA-sim. A gap between Flint and Post-Exec + ASTRA-sim indicates differences in the captured workload graph. A gap between Post-Exec + ASTRA-sim indicates limitations in the cost model itself. This is a factor that can be overcome with higher fidelity, proprietary cost models.

\cref{fig:e2e} shows the comparison result across the two configurations. We observe that the modeled per-iteration duration aligns well across the three configurations.

\section{Evaluation: Case Studies}
\label{sec:eval}
We showcase how \sys{} can guide Design Space Exploration across multiple layers that span from workload optimizations to network hardware.

\subsection{Case study: Operation Reordering in FSDP across Model and Hardware Configuration}
\label{subsec:usecase_fsdp}
\insertFigure{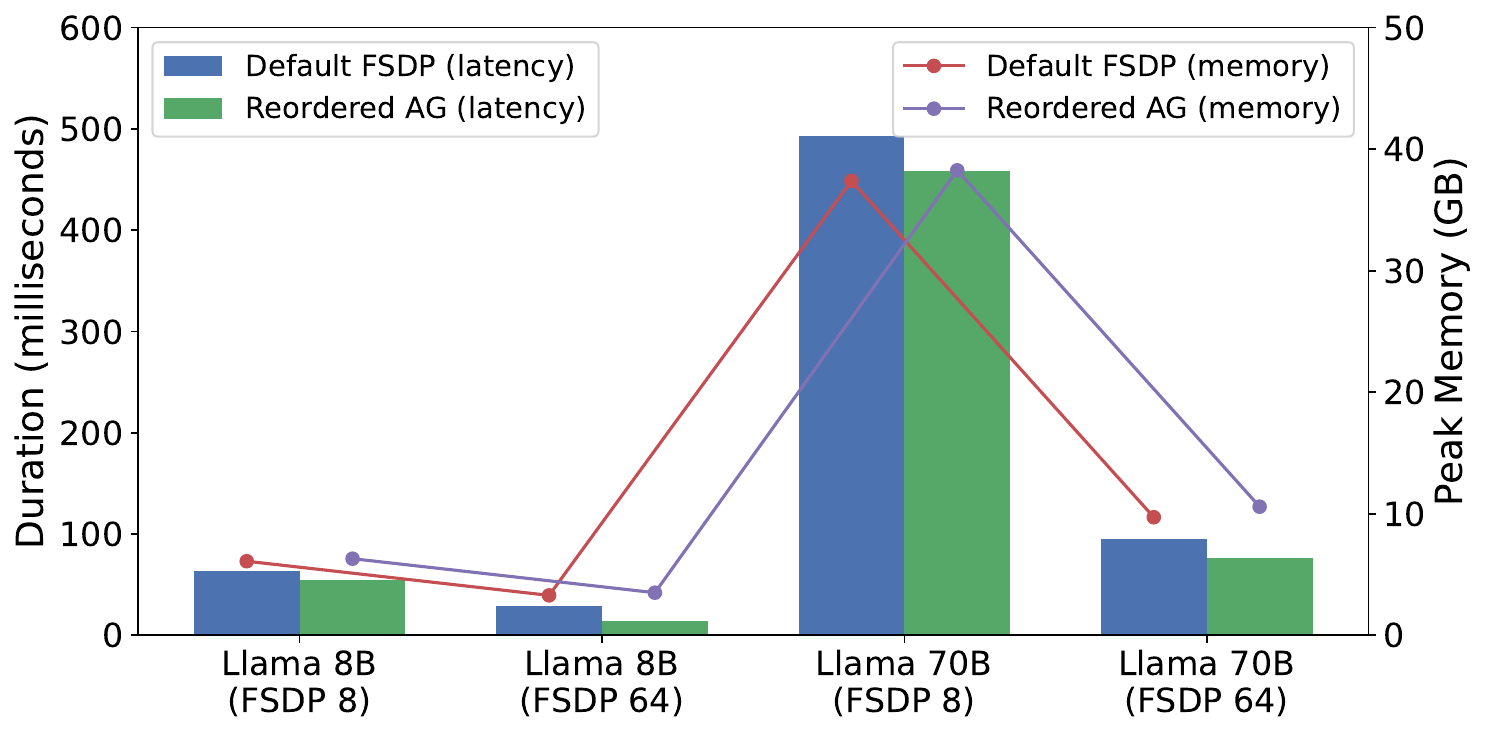}{Per-iteration duration and memory tradeoff of communication reordering in FSDP across scale and model size.}{1}{-2em}{0em}
\insertFigure{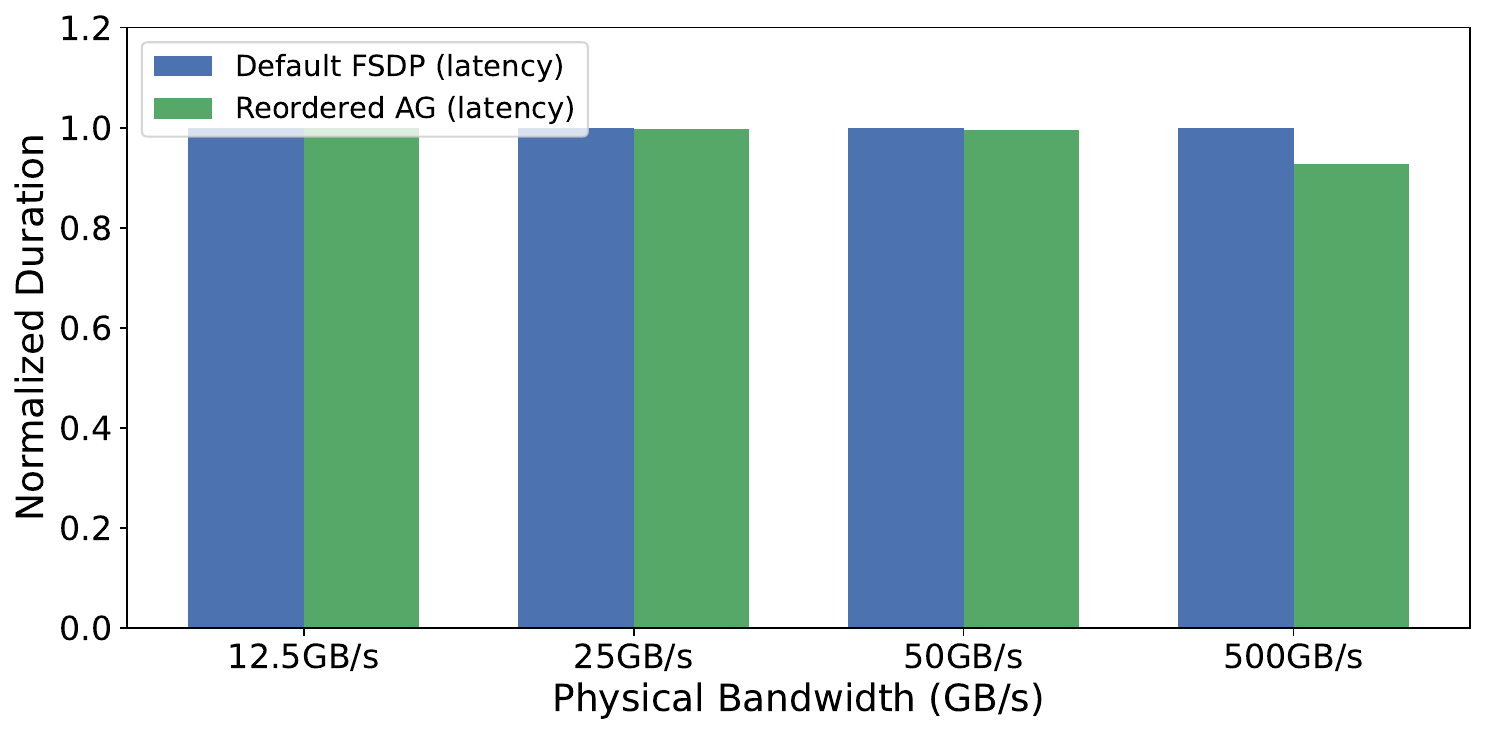}{Per-iteration duration comparison of Reordered AllGather across different interconnect bandwidth. Here Llama 70B model is used.}{1}{-2em}{0em}

We evaluate the scheduling choice of communication reordering in FSDP discussed in~\cref{subsec:graph}. Instead of limiting ourselves to simply comparing the scheduling decision while \textit{fixing} other aspects, we also explore how this scheduling interacts with the model characteristics or the physical topology. 

\cref{fig:fsdp_workload} shows the memory and duration tradeoff of the reordering scheme across different model sizes and parallelization degrees. For each configuration, as outlined in the X axis, we generate two sets of Chakra graphs using \sys{}. One graph is generated after running a graph modification pass on the FX Graph to reorder the collective nodes. 

When running the Llama 8B model across 64 ranks, we get the largest duration benefit of 50\% reduction at a cost of 6.7\% or 0.22GB increase of memory consumption. Even with the larger Llama 70B model, the reordering reduces the duration by 7\% at a marginal memory cost of 2.32\%, or 0.87GB at 8 ranks.

We now explore how the scheduling decision affects the workload behavior across different hardware configurations, specifically the bandwidth of the interconnect network. Here, we take the Chakra graph for the Llama 70B model across 8 GPUs, but run it through interconnects of varying bandwidth. \cref{fig:fsdp_topo} shows the normalized duration comparison across the bandwidth setting. We normalize the value because the per iteration duration increases exponentially as we reduce the available bandwidth. Also note that because the workload and GPU device is the same across configurations, the peak memory stays the same and hence is not depicted here. 

While reordering reduces the duration by 7\% in a high bandwidth scenario, there is marginal difference at lower bandwidth scenarios. This is because at lower bandwidths the communication duration far outweighs the computation duration. For the reordering scheme to work there must be both exposed communication and computation that can be hidden by reordering compute and communication to overlap with each other. However, because the communication is disproportionately large in low bandwidth configurations there is little room for performance improvement.

Note how this case study showcases the power of \sys{}. We are able to study \textit{what-if} scenarios on clusters beyond the number of GPUs readily available to us. Leveraging the compiler IR preserves the data dependency between operations, allowing us to explore scheduling decisions without fear of breaking the model behavior.

\subsection{Custom Collectives in Wafer Scale Compute}

\insertSubFigs{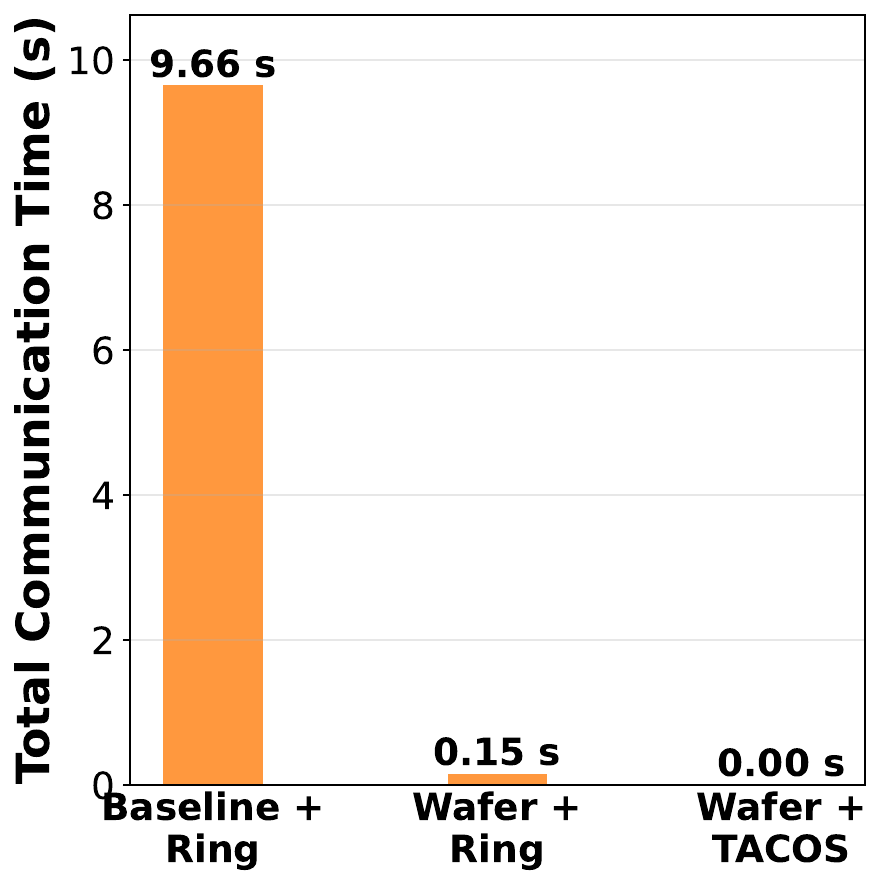}{Total Communication Time}{wafer-comm}{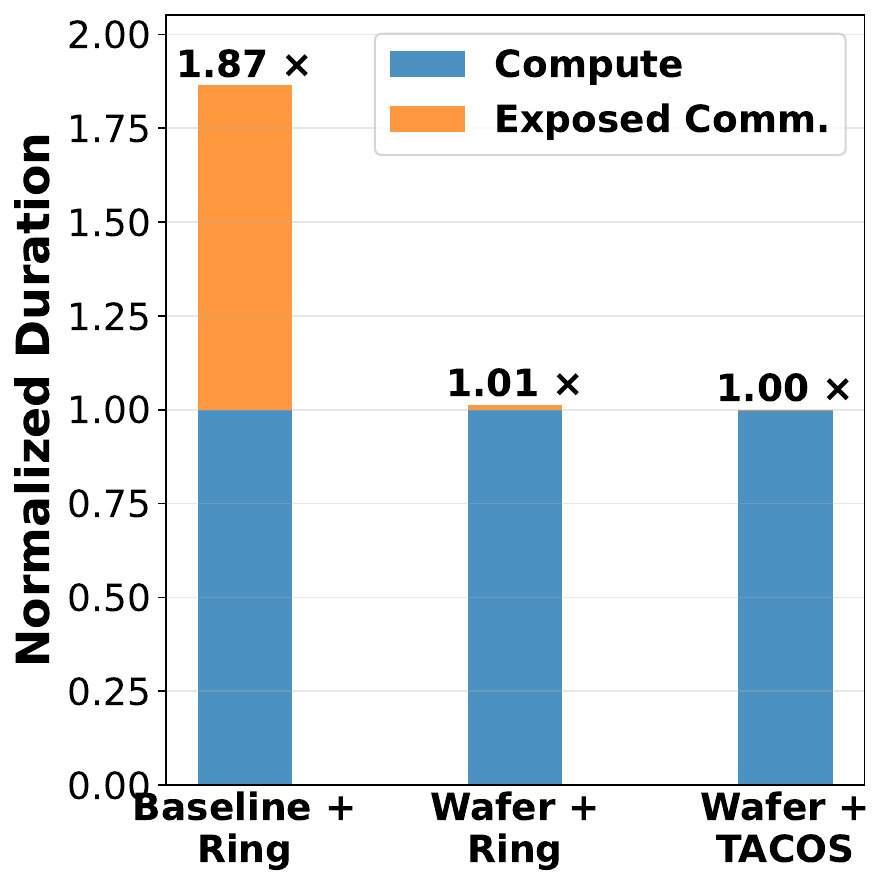}{Normalized Runtime}{wafer-scale}{Simulation results of running Llama3 70B model on a wafer-scale package v. traditional switch topology.}{wafer-fig}{-1 em}

Novel hardware technologies, such as wafer-scale compute or optical network, strive to alleviate the communication bottleneck. In wafer-scale compute, multiple GPUs are placed on the same package in a 2D layout. While the new technology will improve the communication speed, one could ask if traditional Ring based collective algorithms are sufficient. Synthesized collective algorithms, specifically tuned to the 2D Mesh like topology of wafer-scale packages, could potentially further improve performance\cite{fred, cerebras}.

This is a usecase where \sys{} can help in multiple aspects. As the key focus of this usecase is on a hard-to-access network, obtaining post-execution traces becomes less important. While it is difficult to implement arbitrary collective algorithms in a simulator, or even in NCCL, prior art extended ASTRA-sim to simulate custom collective algorithms represented in a separate Chakra graph consisting of point-to-point messages, and displayed how this could be used to study the effect of custom algorithms on end to end workload. \sys{} can leverage this work easily because it already uses Chakra graphs to represent workload information. 

We compare the traditional Ring algorithm with TACOS, a topology aware collective synthesis tool~\cite{tacos}. We simulate the TACOS-generated algorithms by representing them in separate Chakra graphs consisting of point-to-point messages and feeding them into the simulator. We use Llama3 70B model with FSDP=16 as the workload.

\cref{fig:wafer-scale} shows the sum of all communication for the three configurations. The communication decreases by 62$\times$ from Baseline to Wafer + Ring, due to a change in technology. Between Wafer + Ring and Wafer + TACOS, the topology aware collective reduces communication by 51$\times$ by avoiding congestion on the 2D mesh topology. However, the end to end performance improvement shown in \cref{fig:wafer-scale} is rather limited. This is because beyond a certain point, the communication is no longer the dominant bottleneck, and hence there is a diminishing return of performance improvement. We anticipate that the impact of collective algorithm on the workload performance could change for different model and hardware configurations.

\subsection{Physical Network Testing with Network Emulator}

\insertFigure{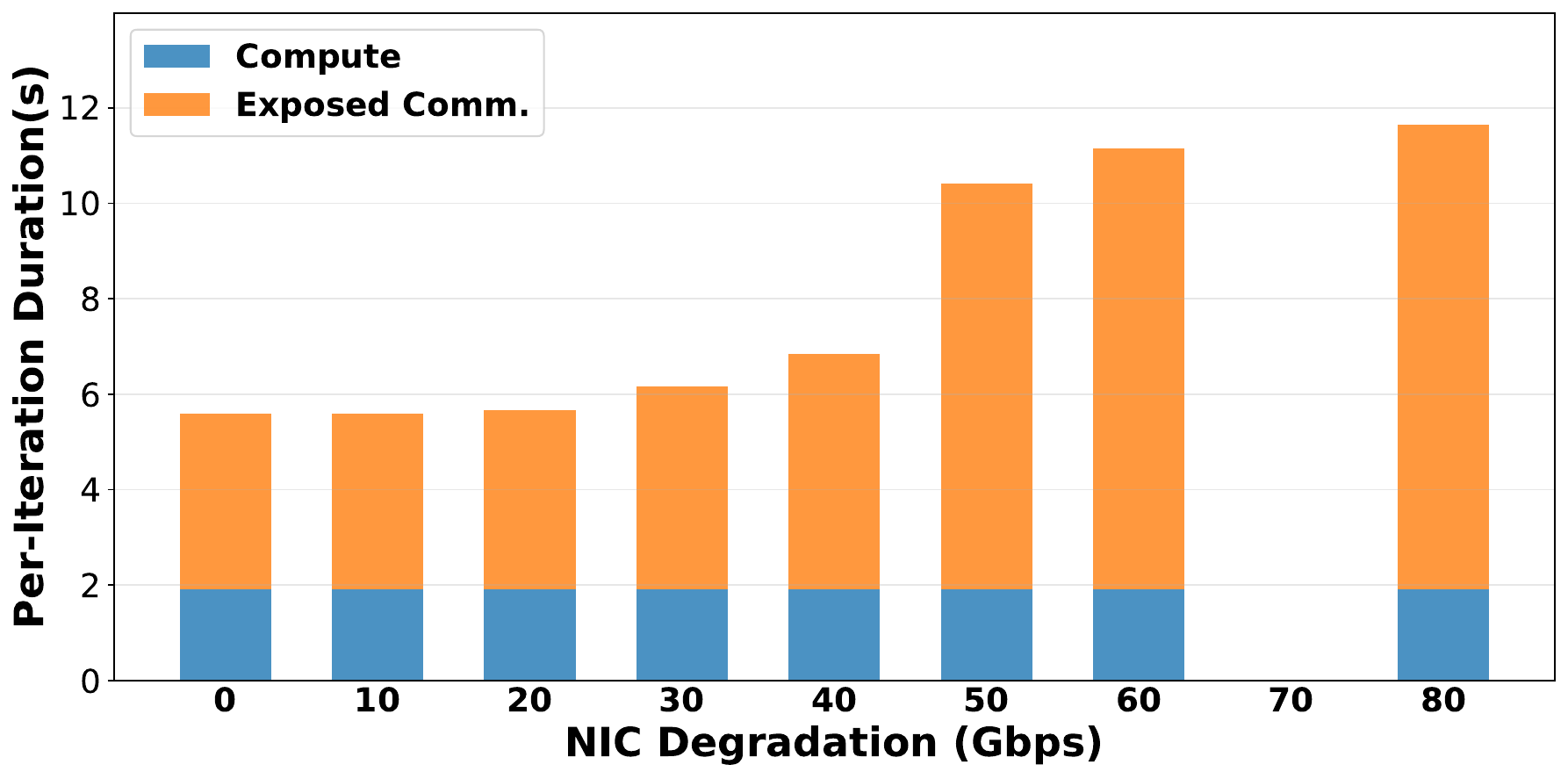}{Per-iteration duration across different NIC degradation. 70Gbps is blank because perftest does not support that rate limit}{1}{-2em}{0em}

In alleviating the communication bottleneck, it is important to understand, configure, verify, and diagnose the underlying network fabric. While network simulators can model the fabric to an extent, some jobs need to be done directly with the physical NIC, such as hardware quality control, tuning features not in simulation, etc~\cite{metaroce, megascale}. 

Genie proposes a testing tool for scale-out network that generates real RDMA traffic but only with CPU nodes~\cite{genie}. That is, users do not need GPU just for the sake of network testing. It uses Chakra graphs to track the operation and their dependencies (i.e. launch a communication only when the upstream operations finish).

A bottleneck in the vision of Genie is that it is difficult to obtain Chakra graphs. This is yet another usecase where \sys{} can come to the rescue. With \sys{}, developers can generate high fidelity workload graphs without needing GPUs and feed it to a cost model - Genie - to test real network fabric.

One usecase of Genie is to identify faulty hardware (flapping NIC, etc.) before attaching and wasting precious GPU time. Since we cannot physically damage the NICs, we emulate NIC degradation by generating background traffic using \texttt{ib\_write\_bw} at different rate limits. Our workload is Llama3 70B with DP=32. Since each node has only one 100Gbps InfiniBand scale-out network, and our topic of interest is the scale-out network, we run the workload across 32 nodes. \cref{fig:genie} shows how the per-iteration duration worsens as we introduce more degradation (i.e. more background traffic). We can see that \sys{} is able to capture delays in network and show it as an increase in communication time. Before attaching GPUs, developers can simply run Genie with the workload graph from \sys{} to ensure that the network fabric is performing correctly. Note that while we struggled to gain access to a single node with 8 GPUs, getting 32 CPU nodes was almost instantaneous.

\section{Discussion and Future Work}
\label{sec:discussion}
While \sys{} focuses on leveraging FX Graph and PyTorch, we envision that our findings can be extended to other frameworks such as JAX. Similar to how `torch.compile` traces FX graphs from model code, JAX programs can be traced to generate Jaxprs (JAX expressions). Develoeprs can write `custom interpreters' that can take and modify Jaxprs and. Capturing Jaxprs and creating Chakra graphs with the intend of feeding into downstream tools is a promising approach to extend \sys{} beyond PyTorch.

Blindly iterating through the whole search space for a better configuration takes a prohibitively long time to converge, as each iteration of the cost model takes time to return results. There are several works that try to speed this process, either by using Reinforcement Learning or user provided hints to prune the search space. Integrating these methods into \sys{}'s feedback loop enables will further enhance the benefit that \sys{} provides. 

One possible future work is to leverage \sys{} to reconfigure the runtime setting, especially in multi tenant environment. Previously, two jobs sharing a cluster only search through the subsection of resources that they start with. However, \sys{} allows jobs to search for optimal settings beyond the deployed cluster. For example, \sys{}'s lack of being constrained to the resource can allow it to be used to adjust the resource split between the two jobs. 

\sys{} can leverage \texttt{torch.compile} to trace code written in the standard \texttt{aten} or \texttt{c10d} APIs. However, \texttt{torch.compile} does not work well with data dependent control flow or custom functions resulting in graph breaks and gaps in the generated workload graph. However, the coverage of \texttt{torch.compile} is continuously increasing, and frameworks such as Torchtitan continuously add more models that can be traced natively with \texttt{torch.compile}.

Because it leverages model code as-is, \sys{} requires relatively less burden in maintaining the codebase and following the rapid advancements in  the field. While it does make changes to the PyTorch API, we anticipate that the PyTorch API will shift less frequently compared to changes in the high level model code itself.

\section{Related Work}
\label{sec:related}
\subsection{Workload graph generation}

There are several work that generate workload graphs at different stage of the model execution. With some effort these graphs could be converted to each other to some extent. Therefore, we focus on \textit{how} these graphs are generated, rather than \textit{what} the schema is.

\textbf{Post-execution Trace:} Chakra ET combines metrics from Kineto traces and dependency information from PyTorch Execution Observer to generate a DAG that encodes both operator information and dependency~\cite{chakra,mystique}. ATLAHS traces NCCL kernel launches for communication and designates the time gap between NCCL operations as a block of compute~\cite{atlahs}. Because it simply merges multiple operations into one timed blob, this trace format does not encode any operator information. While this helps in studying communication, it cannot be used to study different compute operations or different hardware settings. Both approaches need a physical cluster, which is easily prohibitive.

\textbf{Pre-execution Trace from Symbolic Representation:} STAGE takes the symbolic representation of a model and its parallelization, and synthetically generates graphs using the Chakra schema~\cite{stage}. Unity proposes its own graph schema, which also encodes both compute and communication operators in a DAG~\cite{unity}. Similarly, Calculon takes a description of the workload and directly models the performance using its internal analytical tool~\cite{calculon}. SimAI also uses a text based representation to generate graphs~\cite{simai}. Both SimAI and ATLAHS expands the collective operations to point-to-point send and receive messages. However, this expansion is hardcoded and limited to popular algorithms such as Ring or Tree. Other similar approaches lack the ability to model arbitrary model code and optimizations, or work across multiple cost models~\cite{oobleck, mist, pipedream, dapple}.

\textbf{Compiler Internal Representations:}
Deep Learning frameworks leverage pre-execution graph IR for compilation. PyTorch generates FX graphs, while JAX generates Jaxpr, which is later lowered into HLO that the XLA compiler undestands~\cite{fx,jax}. These runtime compilers need to be deployed on real systems to be able to collect and optimize these graphs.

\subsection{Consuming Workload Graphs}
\textbf{Cost Models and Design Space Exploration: }
Several work such as runtime compilers leverage the DL frameworks' IR as pre-execution graph to search for optimizations~\cite{gspmd, whale}. Inductor is the default compiler backend for PyTorch and performs both graph optimization such as scheduling and fusion, and code generation through OpenAI Triton~\cite{pytorch2}. Unity uses the PCG graph to optimize workload parallelization and operator fusion, while Alpa takes Jaxpr and HLO and searches for an optimal parallelization strategy~\cite{unity,alpa}. None of the above works leverage the pre-execution graph to search beyond compiler optimization, such as model architectures or collective optimization. By collecting IR in a GPU-less environment, \sys{} lifts the restriction that compiler IR must be collected and studied within a runtime deployment.

\textbf{Several Use Cases of Chakra Graph: }
Chakra ET has several downstream tools, which collectively help improve our understanding of distributed ML. The ASTRA-sim simulator has a modular design, where users model each Chakra node with varying fidelity based on their problem~\cite{astrasim-2.0}. The PARAM benchmark tool allows for replay of Chakra ET on a real cluster~\cite{mystique}. Genie generates real network traffic from a CPU-only cluster following the workload behavior captured in Chakra graphs~\cite{genie}. This allows users to study the behavior of the real network without needing any GPU. While this work focuses on the ASTRA-sim simulator, users can feed the graph that \sys{} generates into these works.

\section{Conclusion}
\label{sec:conclusion}

We present \sys{}, a tool for obtaining accurate workload representation for GPU-less design space exploration. \sys{} leverages compiler IR to capture the workload behavior with minimal source code modification. \sys{} works under the philosophy of granting the most freedom in exploring the design space, while capturing details that are beyond the scope of design space exploration, such as the internals of the PyTorch framework. We build and showcase an end-to-end workflow, identifying and applying the necessary changes to execute \sys{} on a GPU-less environment.

\bibliographystyle{ACM-Reference-Format}
\bibliography{references}

\end{document}